\documentclass[aps,prd,12pt,citeautoscript,superscriptaddress,showpacs]{revtex4-1}
\usepackage{xcolor}
\usepackage[colorlinks=true, pdfstartview=FitV,linkcolor=blue, citecolor=blue, urlcolor=blue]{hyperref}

\newcommand{\be}{\begin{equation}}
\newcommand{\ee}{\end{equation}}
\newcommand{\bea}{\begin{eqnarray}}
\newcommand{\eea}{\end{eqnarray}}

\usepackage{psfrag}
\usepackage{epsfig}
\usepackage{latexsym,array,theorem,mathrsfs,subfigure,bm,float}
\usepackage{amsmath}
\usepackage{amsfonts}
\usepackage{amssymb}
\usepackage{bbm}
\usepackage{color}

\begin{document}

\title{Zero mass limit of Kerr spacetime is a wormhole}

\author{Gary W. Gibbons}
\email{\tt gwg1@cam.ac.uk}

\affiliation{DAMTP, University of Cambridge, Wilberforce Road, Cambridge CB3 0WA, UK}

\affiliation{
Laboratoire de Math\'{e}matiques et Physique Th\'{e}orique CNRS-UMR 7350, \\ 
Universit\'{e} de Tours, Parc de Grandmont, 37200 Tours, France}

\author{Mikhail~S.~Volkov}
\email{\tt volkov@lmpt.univ-tours.fr}
\affiliation{
Laboratoire de Math\'{e}matiques et Physique Th\'{e}orique CNRS-UMR 7350, \\ 
Universit\'{e} de Tours, Parc de Grandmont, 37200 Tours, France}

\affiliation{
Department of General Relativity and Gravitation, Institute of Physics,\\
Kazan Federal University, Kremlevskaya street 18, 420008 Kazan, Russia
}

\begin{abstract} 
\vspace{1 cm}

We show that, contrary to what is usually claimed in the literature, 
the zero mass limit of Kerr spacetime is not flat Minkowski space but a spacetime whose geometry 
is only locally flat. 
This limiting spacetime, as the Kerr spacetime itself, 
contains two asymptotic regions and hence cannot be topologically trivial. 
It  also contains  a curvature singularity, because 
the power-law  singularity of the Weyl tensor vanishes in the limit but there remains
a distributional  contribution  of the Ricci tensor. 
This spacetime can be interpreted as a wormhole 
sourced by a negative tension ring. We also extend 
the discussion to similarly interpret the zero mass limit
of the Kerr-(anti)-de Sitter spacetime.

\end{abstract} 

\maketitle

%\newpage

%\thispagestyle{empty}
%\tableofcontents
%\newpage
%\setcounter{page}{1}
%\setcounter{footnote}{0}

%\bibliographystyle{utphys}
%\addcontentsline{toc}{section}{References}
%\bibliography{blackholes}
\section{Introduction}

In the recent work \cite{Gibbons:2016bok,Gibbons:2017jzk} 
we studied wormholes obtainable from vacuum Weyl metrics via duality rotations.
One of our findings were wormholes
 with locally flat geometry  described by the line element \eqref{0} below (where $r\in(-\infty,\infty)$)
 and sourced by singular rings of negative tension. 
 Such solutions
can  be viewed as a particular case of  the  ``loop based wormholes"  obtained by 
surgeries and identifications performed on Minkowski space \cite{Visser:1995cc}. In this  note we show 
that the same solutions can also be obtained 
by taking the zero mass limit of the Kerr spacetime. 

It is usually  argued in the literature (see for example \cite{landau1975classical,Visser:2007fj}) 
that taking the Kerr black hole mass $M$ to zero reduces the 
curvature to zero hence yielding flat Minkowski space. 
Indeed, the Kerr geometry becomes locally flat in this limit. 
However, it cannot be globally flat and  topologically trivial  because it inherits from the original 
Kerr geometry the non-trivial topology with two asymptotic regions. 
At the technical level, this  means that  the radial coordinate 
in the metric  spans a line and not a half-line.  
It follows that, although the Weyl part of the curvature vanishes when $M\to 0$, 
the curvature also contains a distributional  Ricci part 
supported by the ring  which does not vanish in the limit.
This can be interpreted as an effect of the matter source -- a ring made of a cosmic string 
with a negative tension.  
The geometry outside the ring is locally flat and has two asymptotic regions connected by a throat --
the disk encircled  by the ring. 

In what follows we first describe how the locally flat wormholes can be obtained via analytic 
continuation and then discuss the relation to the Kerr metric. 

\section{Flat space in oblate spheroidal  coordinates \label{I}}  
Let us consider Minkowski  metric,
\bea                      \label{1}
ds^2&=&-dt^2+dx^2+dy^2+dz^2,
\eea
and pass first to the cylindrical coordinates with 
$x=\rho\cos\varphi$, 
$y=\rho\sin\varphi$,  so that 
\be                      \label{1a}
ds^2=-dt^2+d\rho^2+\rho^2d\varphi+dz^2,
\ee
and then further transform to the oblate spheroidal coordinates by setting 
\be                  \label{c}
\rho=\sqrt{r^2+a^2}\sin\vartheta,~~~~~z=r\,\cos\vartheta, 
\ee
where $r\in [0,+\infty)$, $\vartheta\in[0,\pi]$. This  gives 
\be                       \label{0}
ds^2=-dt^2+\left.\left.\frac{r^2+a^2\cos^2\vartheta}{r^2+a^2}\right[dr^2+(r^2+a^2)d\vartheta^2\right]
+(r^2+a^2)\sin^2\vartheta \,d\varphi^2.
\ee
The Jacobean of the coordinate transformation 
\be
\left|\frac{{\cal D}(x,y,z)}{{\cal D}(r,\vartheta,\varphi)}\right|=(r^2+a^2\cos^2\vartheta)\sin\vartheta
\ee
vanishes
at $r=\cos\vartheta=0$, hence for 
\be                       \label{ring}
\rho^2\equiv  x^2+y^2=a^2,~~~~~z=0. 
\ee
This corresponds to 
a ring  of radius $a$ in the equatorial plane. As a result, 
the $(r,\vartheta,\varphi)$ coordinates cover the whole of Minkowski space, excluding the $z$-axis and 
 the ring. The coordinate singularity on the axis  may be treated in the 
 standard way and we shall  discuss it no further. 
 Let us consider the coordinate singularity at the ring. 
One has 
\be
\frac{\rho^2}{r^2+a^2}+\frac{z^2}{r^2}=1,
\ee
hence lines of constant $r$ are oblate (half)-ellipses in the $(\rho,z)$ plane, 
orthogonal to them are hyperbolas of constant $\vartheta$ (see Fig.\ref{Fig}). 
In the $r\to 0$ limit the ellipses shrink to the segment of the $\rho$-axis, 
\be                       \label{seg}
{\cal I}=\{\rho\in [0,a],z=0\},
\ee
whereas the $\vartheta$ coordinate is discontinuous across  the segment since  
\be
\lim_{z\to \pm 0}\cos\vartheta=\pm \sqrt{1-\rho^2/a^2}~~~\mbox{if}~~~\rho\leq a,~~~~~
\lim_{z\to 0}\cos\vartheta=0~~~\mbox{if}~~~\rho\geq a.
\ee
This can be understood as follows.
 The inverse coordinate transformation $(\rho,z)\to(r,\vartheta)$,
\be                      \label{inv}
r+ia\cos\vartheta=+\sqrt{\rho^2+(z+ia)^2},
\ee
has a branch point at $(\rho,z)=(a,0)$ and the segment \eqref{seg} corresponds to 
the branch cut position.  
Choosing only one branch of the square root, its real part $r$  is non-negative but  the 
imaginary part $a\cos\vartheta$   is then necessarily discontinuous across the cut. 

As a result, the $(r,\vartheta)$ coordinates are discontinuous  at the disk 
of radius $a$ in the equatorial plane. A timelike geodesic along the $z$-axis,
which is simply a straight line in the $(x,y,z)$ coordinates, is described 
in the $(r,\vartheta)$  coordinates by 
\be                            \label{geod1}
\frac{d r}{ds}=\pm \sqrt{{\cal E}^2-\mu^2},~~~~\frac{d \vartheta}{ds}\sim \sin\vartheta=0,
\ee
where ${\cal E},\mu,s$ are the particle energy, mass, and proper time, respectively. 
Since $r$ should be non-negative, 
one is bound to choose opposite signs in front of $\sqrt{E^2-\mu^2}$ and also 
different values of $\vartheta$ (either $0$ or $\pi$ for this geodesic)  at the opposite  sides of the disk. 
As a result, $r(s)$ is not smooth while $\vartheta(s)$ is discontinuous across the disk.

\section{Wormhole via analytic extension   \label{II}}

The metric \eqref{0}  can be 
geodesically extended to negative values of $r$.
Indeed, if $r$ is allowed to become negative, then 
there is no need to change sign in front of $\sqrt{E^2-\mu^2}$ in \eqref{geod1} across the disc, hence 
$r(s)$ is smooth. As we shall see in a moment, 
there is no need either to require that $\vartheta(s)$ jumps. 
Therefore, the geodesics analytically continue from $r>0$ to the $r<0$ region. This
applies not only to geodesics along the $z$-axis but to all geodesics 
{which do not hit the ring} $(\rho,z)=(a,0)$. 
As a result, the metric  in \eqref{0} naturally extends to $r\in(-\infty,+\infty)$. 

\begin{figure}[th]
\hbox to \linewidth{ \hss

	\resizebox{13cm}{9cm}{\includegraphics{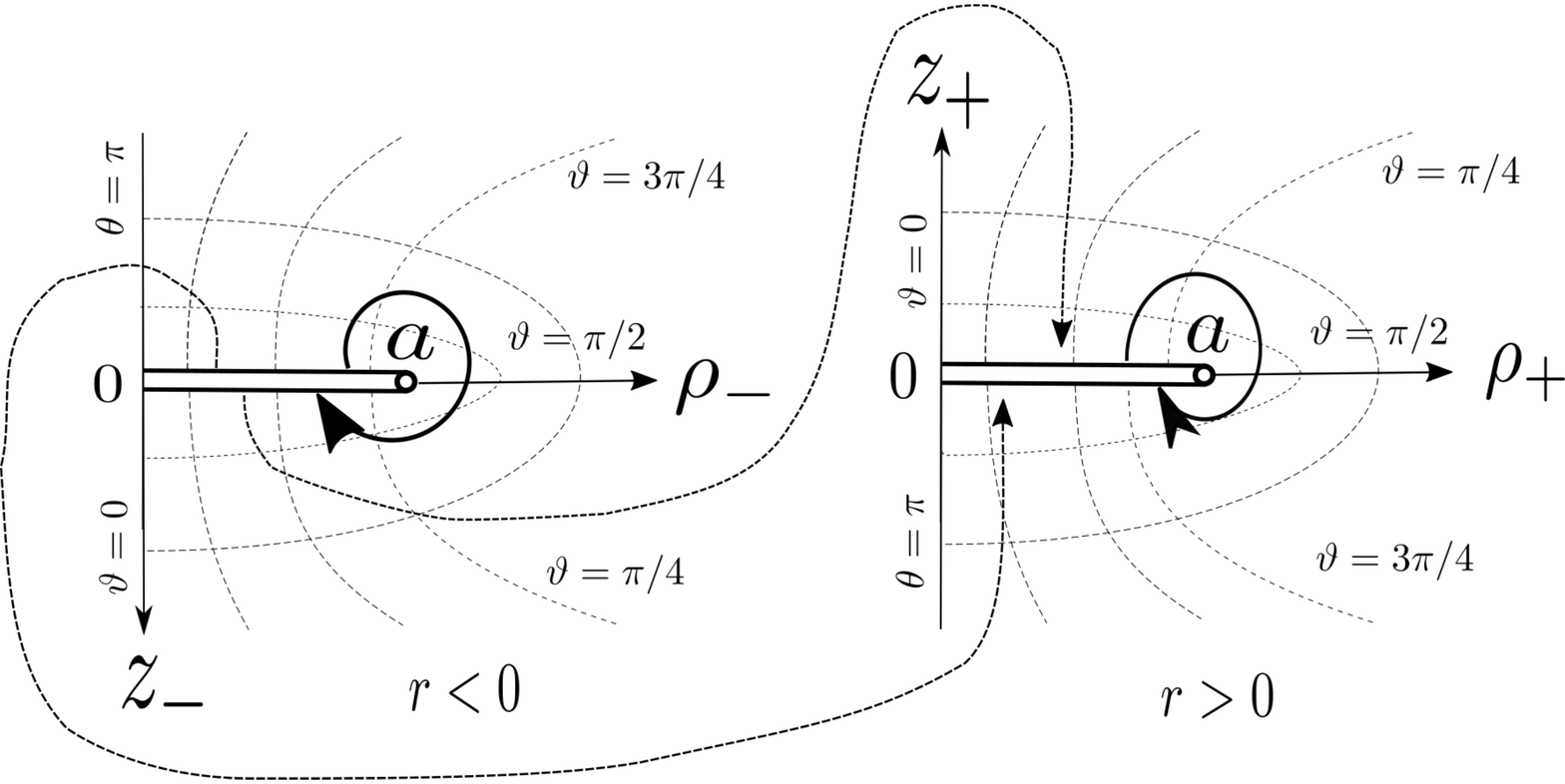}}
\hspace{1mm}
	
\hspace{1mm}
\hss}
\caption{Two  charts $(\rho_{+},z_{+})$ and $(\rho_{-},z_{-})$
 needed to cover the geometry \eqref{0} with $r\in(-\infty,\infty)$. Each chart has a branch 
cut along the $[0,a]$ segment of the $\rho$-axis.  Lines of constant $r$ are oblate (half)-ellipses, 
orthogonal to them are hyperbolas  of constant $\vartheta$. 
The $r,\vartheta$ coordinates are discontinuous through the cut 
on each individual chart, but they smoothly continue from one chart to
the other if  the upper edge of one cut is glued to the lower edge of the other and vice versa
as shown. 
A contour around the branch point 
$(a,0)$ then performs  one revolution in the $(\rho_{+},z_{+})$  chart,
followed by a second revolution in the $(\rho_{-},z_{-})$ chart, and only after  that  closes. 
 }
\label{Fig}
\end{figure}

When expressed in the $(\rho,z)$ coordinates, 
the metric is still manifestly flat and is given by  \eqref{1a},  but the speciality now  is that 
 the coordinate transformation \eqref{c} is no longer  bijective. Since
$(r,\vartheta)$ and $(-r,\pi-\vartheta)$ map to the the same $(\rho,z)$, it follows that 
when $r$ and $\vartheta$ span all their values, $\rho$ and $z$  will span all their values twice.
Therefore,  one needs two  $(\rho,z)$ charts to cover the spacetime, let us call them 
$(\rho_{+},z_{+})$ and $(\rho_{-},z_{-})$. 
In each  chart the metric has the form \eqref{1a}, but one has 
\be                      \label{inv1}
r+ia\cos\vartheta=\pm\sqrt{\rho_\pm^2+(z_\pm+ia)^2}
\ee
hence one chart  spans the Riemann sheet where $r>0$,
the other chart spans the sheet where $r<0$,
  while  together they span 
the whole of the Riemann  surface. 
Each Riemann sheet has a  branch cut along the segment \eqref{seg} and the two sheets are 
glued to each other  along the cuts by identifying the upper side of one cut with the lower side of the other 
and vice versa (see Fig.\ref{Fig}). 
The $\vartheta$-coordinate then changes continuously when passing from one chart 
to the other while the $r$-coordinate simply  passes through zero and changes sign.  

As a result, the spacetime actually consistes of two copies of $R^4$ 
glued together through the disk, hence this is a wormhole with 
two asymptotic regions. The geometry is locally flat and the curvature is 
locally zero, but not globally since the ring at $(\rho_\pm,z_\pm)=(a,0)$ now supports a physical 
singularity of the curvature. To see this one notices that a contour around the branch point $(a,0)$ 
in the $(\rho_+,z_+)$ chart does not close after a revolution of $2\pi$ but continues to the $(\rho_-,z_-)$ chart
and only after a second revolution of $2\pi$ returns to the original chart to close. As a result, the total 
angle increment is $4\pi$, therefore there is a {\it negative} angle deficit of $2\pi-4\pi=-2\pi$
and hence the conical singularity of the curvature at the  ring. 

One arrives at the same conclusion using the $(r,\vartheta)$ coordinates. Introducing 
$x_1=r/a$ and $x_2=\cos\vartheta$, the metric \eqref{0} reduces for small $x_1,x_2$ to 
\be                       \label{000}
ds^2=-dt^2+(x_1^2+x_2^2)[dx_1^2+dx_2^2]+ a^2 d\varphi^2 
=-dt^2+dx^2+x^2d\theta^2+ a^2 d\varphi^2, 
\ee
where $(x_1+ix_2)^2=(2/a)\,x\exp\{i\theta\}$.  Since $\theta\in[0,4\pi)$, the metric contains a conical
singularity at $x=0$ stretching along the azimuthal $\varphi$-direction 
(see \cite{sokolov1977} for an account of ``bent" conical singularities).
This curvature singularity can be interpreted 
as that corresponding to a matter source -- 
 a ring made of infinitely thin cosmic string with negative tension 
 \cite{Gibbons:2016bok,Gibbons:2017jzk} 
\be                         \label{T}
T=-\frac{c^4}{4G}. 
\ee
The ring ``cuts a hole" in spacetime and acts as 
a ``gate" connecting the $r>0$ universe and the  $r<0$ universe.
To create  a ring wormhole of one metre radius one needs a negative energy equivalent to the 
mass of Jupiter  
(see \cite{Gibbons:2016bok,Gibbons:2017jzk}  for further details).

\section{Zero mass limit of Kerr spacetime}

Summarising the above discussion, depending on the choice of  range of the radial coordinate,
the same metric \eqref{0} describes either flat 
Minkowski space or the wormhole with locally flat geometry. 
Let us now see that the latter 
can also  be obtained from the Kerr spacetime by taking the black hole mass to zero. 

Consider  the Kerr metric \cite{Kerr:1963ud} expressed in  Boyer-Lindquist coordinates \cite{Boyer:1966qh},
\bea                          \label{Kerr}
ds^2&=&-dt^2+\left.\left.\frac{2Mr}{\Sigma}\right(dt-a\sin^2\vartheta\, d\varphi\right)^2
+\Sigma\left(\frac{dr^2}{\Delta}+d\vartheta^2\right)+(r^2+a^2)\sin^2\vartheta d\varphi^2\,; \nonumber \\
\Delta&=&r^2-2Mr+a^2,~~~~~\Sigma=r^2+a^2\cos^2\vartheta.
\eea
As is well known \cite{Carter:1968rr}, 
the radial coordinate here can be both positive or negative, 
$r\in(-\infty,\infty)$, and there are 
two asymptotic regions corresponding to $r\to\pm\infty$ 
 (see \cite{Adamo:2014baa,Heinicke:2015iva,Teukolsky:2014vca}
for recent reviews of the Kerr metric). 
The black hole mass has opposite signs 
when viewed from these two regions.
The curvature invariant 
\be         \label{curv}
R_{\mu\nu\alpha\beta}R^{\mu\nu\alpha\beta}=
C_{\mu\nu\alpha\beta}C^{\mu\nu\alpha\beta}
=\frac{48 M^2(2r^2-\Sigma)(\Sigma^2-16r^2a^2\cos^2\vartheta)}{\Sigma^6}
\ee
diverges at $\Sigma=0$, that is at $r=\cos\vartheta=0$, which corresponds to 
a ring in the equatorial  plane. 
The singularity is shielded by the horizon if $M^2>a^2$ and is naked if $M^2<a^2$. 

\begin{figure}[th]
\hbox to \linewidth{ \hss

	\resizebox{8cm}{6cm}{\includegraphics{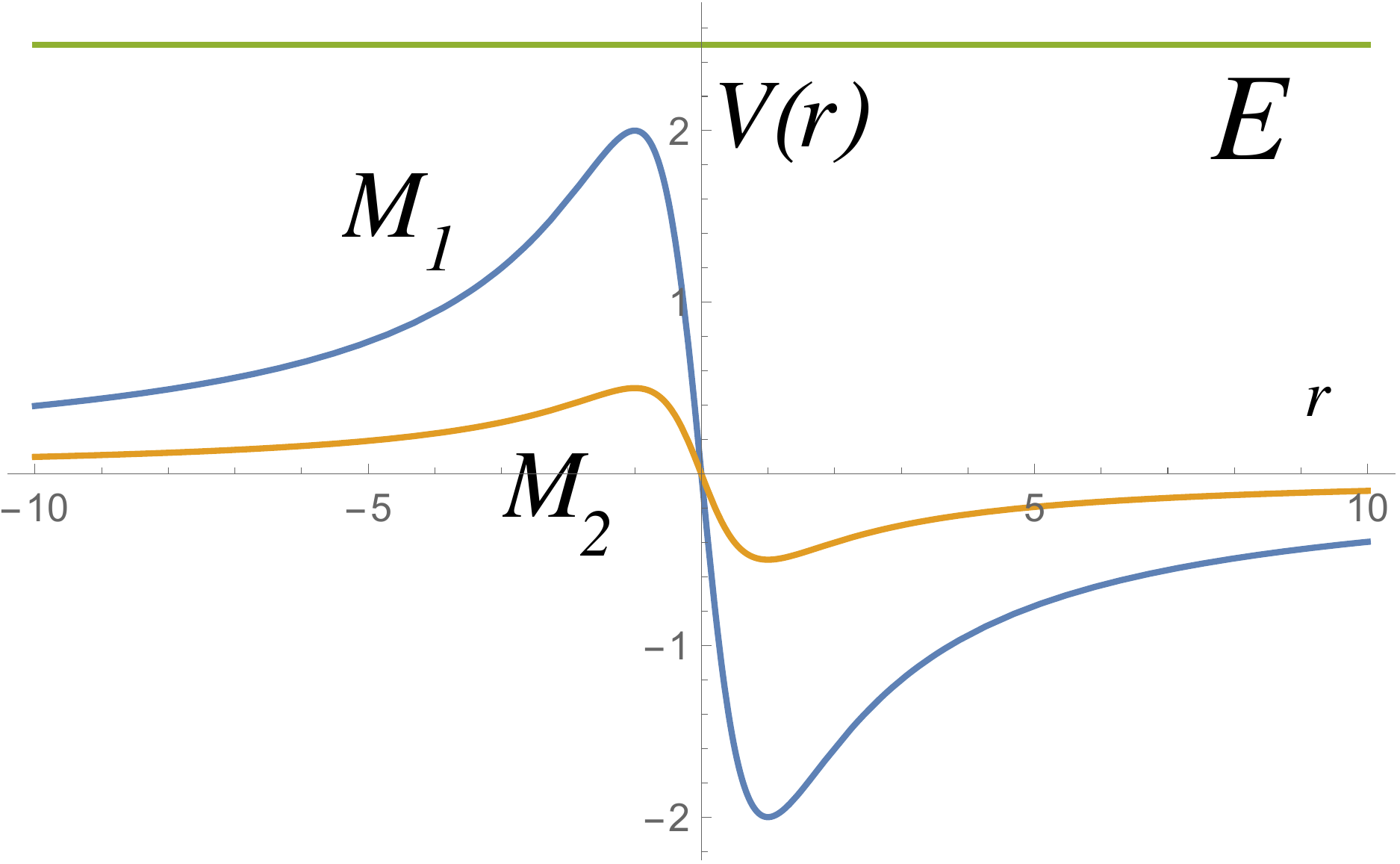}}

\hss}
\caption{Potential $V(r)$ in the geodesic equation \eqref{geod} for two values of the black hole mass, $M_1>M_2$.
When $M\to 0$
the potential vanishes letting the particle freely move in the interval $r\in(-\infty,+\infty)$. 
}
\label{Fg1}
\end{figure}

The geodesics which do not belong to the equatorial plane 
miss the ring  singularity  and pass from the $r>0$ region to the $r<0$ region. 
For example, a  timelike geodesic along the symmetry axis is described by 
\be                \label{geod}
\frac{1}{\mu^2}\left(\frac{dr}{d s}\right)^2+V(r)=E~~~~~\mbox{with}~~~~~~V(r)=-\frac{2Mr}{r^2+a^2}
\ee
where $E={\cal E}^2/\mu^2-1$. The potential $V(r)$ is attractive 
for $r>a$ and repulsive for $r<-a$ and is perfectly  regular at $r=0$ (see Fig.\ref{Fg1}). 
If $E$ is larger than the maximal value of the potential,  $V_{\rm max}=M/a$,
then $r(s)$ interpolates over  the whole range,  $r\in(-\infty,+\infty)$. 

Let us fix  $a\neq 0 $ and take the limit
$M\to 0$.  The potential  $V(r)$ then uniformly tends to zero letting the particle move  freely
in the interval  $r\in(-\infty,+\infty)$. 
The Kerr metric \eqref{Kerr} reduces in this limit precisely to \eqref{0}
and describes, since $r\in(-\infty,+\infty)$, 
the locally flat wormhole and not flat Minkowski space as is 
usually assumed in the literature. 

A remark is in order here. As was discussed above, the geometry  \eqref{0} 
is locally flat, but this fact alone is not sufficient to determine its global structure 
and one needs in addition to specify the range of the radial coordinate $r$. 
If one is interested in a local geometry in an open set, for example for $r\in(0,\infty)$, 
then it is correct to say that the $M\to 0$ limit of the Kerr metric is flat. 
However, one is not free to choose  the spacetime topology  when one takes the limit. 
The original Kerr spacetime contains two asymptotic regions and the geodesics 
interpolating between them sweep the total interval, $r\in(-\infty,\infty)$. 
These properties should hold 
also for $M\to 0$, hence the topology is non-trivial in this limit and corresponds to the 
wormhole described above. This spacetime still contains a curvature singularity.

In fact, the existence of curvature singularity for $M\to 0$ was emphasised already 
by Carter in  \cite{Carter:1968rr} 
(between Eqs.(4,5) of that  paper) by saying that 
in the special case where $M$ vanishes ``there must still be a curvature singularity at $\Sigma=0$,
although the metric is then flat everywhere else".
Specifically,  the curvature consists 
of the Weyl part and Ricci part. The Weyl part vanishes as $M\to 0$ as seen from \eqref{curv},
while the Ricci tensor is zero outside the singularity since the metric is vacuum. 
However, the Ricci tensor is still allowed to have a non-zero value { at the singularity}, 
even in the $M\to 0$ limit. 
Indeed, 
as we have seen above, the metric  \eqref{0} has the conical singularity at the ring, hence the Ricci tensor 
has a delta-function structure with the support at the ring. 
This is the descendant of the 
black hole ring singularity for $M\to 0$.

Another way to illustrate the same thing  is to return to a finite value of $M$ and  express  the Kerr 
metric in Kerr-Schild coordinates $T,x,y,z$ \cite{Schild} related to  Boyer-Lindquist coordinates 
$t,r,\theta,\varphi$ in \eqref{Kerr} via
\bea                     \label{KS1}
x=\rho\cos\phi,~~~y=\rho\sin\phi,~~~z=r\cos\vartheta,~~~
T=t+\int\frac{2Mr}{\Delta}\,dr,~~~
\eea
where 
\be                  \label{KS2}
\rho=\sqrt{r^2+a^2},~~~~\phi=\varphi+\int\frac{2Mar}{\Sigma\Delta}\,dr,~~~
\ee
which yields 
\bea                     \label{KS3}
ds^2=&-&dT^2+dx^2+dy^2+dz^2  \nonumber \\
&+&\frac{2Mr^3}{r^4+a^2z^2}\left(
\frac{r(xdx+ydy)}{r^2+a^2}
+\frac{a(ydx-xdy) }{r^2+a^2}
+\frac{z}{r}\,dz
+dT
\right)^2.
\eea
One notices that $(\rho,z)$ in these formulas 
 are related to $(r,\vartheta)$ precisely as in \eqref{c}, 
therefore the discussion 
of Section \ref{II} applies literally. 
It follows that, since $r\in(-\infty,+\infty)$, one needs two Kerr-Schild charts  
$(\rho_{+},z_{+})$ and $(\rho_{-},z_{-})$
to 
cover the manifold. Each chart has a branch cut at $\rho\in[0,a]$, $z=0$, and to analytically continue 
from one chart to the other one identifies the upper side of one cut with the lower side of the other 
and vice versa. These facts are identical to those described in Section \ref{II}, but they are 
described also in the Hawking-Ellis book \cite{Hawking:1973uf} 
discussing in Section 5.6 the structure of the supercritical $(a^2>M^2)$ Kerr spacetime. 
Figure 27 in that book   shows how the two charts are glued through the cuts,
and it  is precisely the same 
as our Fig.\ref{Fig}. 

This shows that the conical singularity is present already for $M\neq 0$. Indeed, a contour 
around the core of the ring singularity passes from one Kerr-Schild charts 
to the other and then back to close,
hence the total angle increment is $4\pi$ 
(the terms in the second line in \eqref{KS3} do not influence 
this result if the contour is vanishingly small).
Therefore, the ring source of the Kerr metric 
supports, apart from 
the power-law singularity of the Weyl part of the curvature, also 
 the distributional singularity of the Ricci part of the curvature.
 The former disappears in the $M\to 0$ limit but the latter remains. 
 {The part of the Kerr source that vanishes for $M\to 0$ was computed in \cite{Israel:1976vc},
 while the non-vanishing part corresponds to the negative tension ring.} 

Finally  we recall  that the Kerr spacetime with $M^2<a^2$ contains closed 
timelike curves (CTC) \cite{Carter:1968rr}. 
This is a consequence of the fact that the $g_{\varphi\varphi}$ component of the metric \eqref{Kerr},
\be
g_{\varphi\varphi}=\left(
r^2+a^2+\frac{2Ma^2 r}{\Sigma^2}\sin^2\vartheta
\right)\sin^2\vartheta,
\ee
becomes negative in the $r<0$ region close to the ring, hence closed orbits of the vector $\partial/\partial\varphi$
become timelike. These CTC's can be deformed to pass through any point of the spacetime  \cite{Carter:1968rr}. 
However, if $M\to 0$ then $g_{\varphi\varphi}$ is positive and the problem does not arise. 

\section{Locally (anti)-de Sitter wormholes}
As an application, we consider  the generalisation of the above analysis for a non-vanishing cosmological constant. 
For $\Lambda\neq 0$  one cannot use the  Weyl formulation originally applied in 
\cite{Gibbons:2016bok,Gibbons:2017jzk} to construct wormholes. However, one can 
consider  the $M\to 0$ limit of the Kerr-(anti)-de Sitter ((A)dS) metric \cite{Cart} expressed 
in oblate spheroidal coordinates similar to those used in \eqref{0} \cite{Akcay:2010vt}, 
\bea                       \label{00}
ds^2&=&-\frac{\Delta_\theta}{\Xi}\,D\, dt^2+\frac{r^2+a^2\cos^2\vartheta}{r^2+a^2}\left[\frac{dr^2}{D}
+\frac{r^2+a^2}{\Delta_\theta}d\vartheta^2\right]+\frac{r^2+a^2}{\Xi}\,\sin^2\vartheta d\varphi^2;  \nonumber \\
D&=&1-\frac{\Lambda r^2}{3},~~~~~ \Delta_\theta=1+\frac{\Lambda a^2}{3}\cos^2\vartheta,~~~~~
\Xi=1+\frac{\Lambda a^2}{3};
\eea
 assuming that $\Xi>0$. 
For $\Lambda\to 0$ this metric reduces  to \eqref{0}. For $\Lambda\neq 0$ 
it is singular at the ring $r=\cos\vartheta=0$, similarly to \eqref{0}. 
The metric is regular everywhere else (away 
from the symmetry axis) if $\Lambda\in(-3/a^2,0]$, while for $\Lambda>0$ it is regular (and static)
only for $r^2<3/\Lambda$. 
The coordinate transformation $(r,\vartheta)\to (R,\Theta)$ with 
\be                      \label{RT}
R^2=\frac{1}{\Xi}\left(r^2\Delta_\theta+a^2\sin^2\vartheta\right),~~~~
R\cos\Theta=r\cos\vartheta
\ee
brings the metric to the standard (A)dS form, 
\be
ds^2=-\left(1-\frac{\Lambda R^2}{3}\right)dt^2+\frac{dR^2}{1-\Lambda R^2/3} +R^2
(d\Theta^2+\sin^2\Theta\, d\varphi^2).
\ee
If the radial coordinate $r$ in \eqref{00} changed in the interval
$[0,\infty)$ then the coordinate transformation \eqref{RT}
would be bijective and the geometry \eqref{00} 
would be globally (A)dS. The ring at $r=\cos\vartheta=0$ would then correspond to a coordinate singularity. 
However,  the geometry  \eqref{00} inherits the global structure of the original 
Kerr-(A)dS geometry, hence $r\in(-\infty,\infty)$. 
As a result, the geometry is only locally (A)dS and interpolates between two  (A)dS  regions
connected  through 
the throat -- the disk encircled by the ring at $r=\cos\vartheta=0$. 
The ring itself supports a curvature singularity 
whose existence is revealed by considering a closed contour in the plane spanned by $x_1=r/a$
and $x_2=\cos\vartheta$. The arguments similar to those used around \eqref{000} show that the winding 
angle increases up to $4\pi$, hence there is a conical singularity of the Ricci tensor. This can be interpreted 
as a cosmic string loop with the negative tension $T=-c^4/(4G)$. 

Summarising, the zero mass limit of the Kerr-(A)dS metric \eqref{00} describes a wormhole 
supported by a negative tension ring whose geometry is locally (A)dS. For $\Lambda=0$ it reduces to 
the locally flat wormhole \eqref{0}. Below we describe some properties 
of the geometry \eqref{00}. 

\subsection{Structure of wormholes with $\Lambda=0$ and $\Lambda<0$}

\begin{figure}[th]
\hbox to \linewidth{ \hss

	\resizebox{16cm}{6cm}{\includegraphics{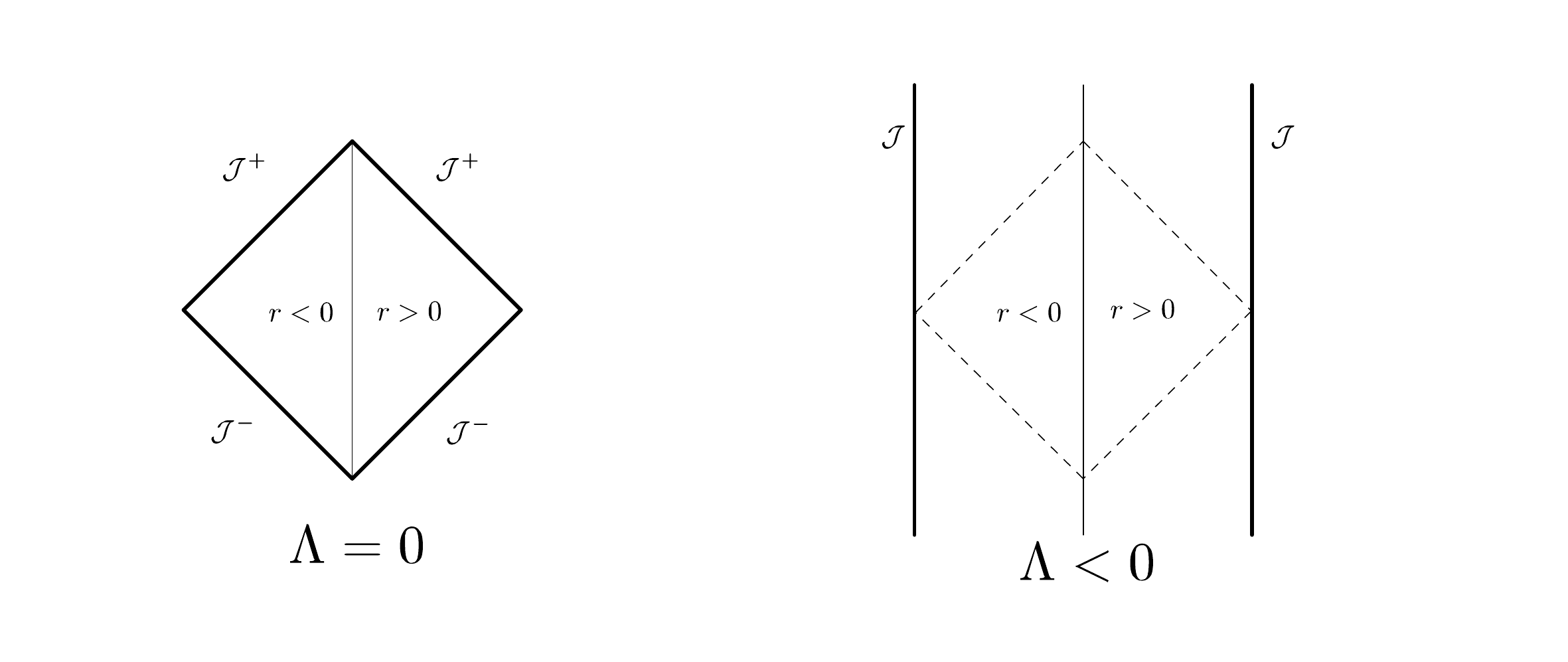}}

\hss}
\caption{Conformal diagrams of wormholes with $\Lambda=0$ and $\Lambda<0$ .
}
\label{Fg3}
\end{figure}

The global  structure of wormholes is simple for $\Lambda=0$ or $\Lambda<0$. They
connect through the disk either two copies of Minkowski space or two copies of AdS space, respectively. 
Considering for simplicity geodesics following the symmetry axes with $\vartheta(s)=0$, 
the conformal diagrams of subspaces spanned by these geodesics  are shown in Fig.\ref{Fg3}. 
The diagram of the $\Lambda=0$ solution is  made of two copies of the Minkowski space 
conformal diagram, one for $r>0$ and the other for $r<0$. The copies are joined across the history of 
the disk at $r=0$ (the disk is represented by one point, $r=\vartheta=0$). Similarly, the diagram 
for the AdS wormhole consists of two copies of the  AdS diagram with the timelike boundary
${\cal J}$.

\subsection{Structure of wormhole with  $\Lambda>0$}
 \begin{figure}[th]
\hbox to \linewidth{ \hss

	\resizebox{20cm}{8cm}{\includegraphics{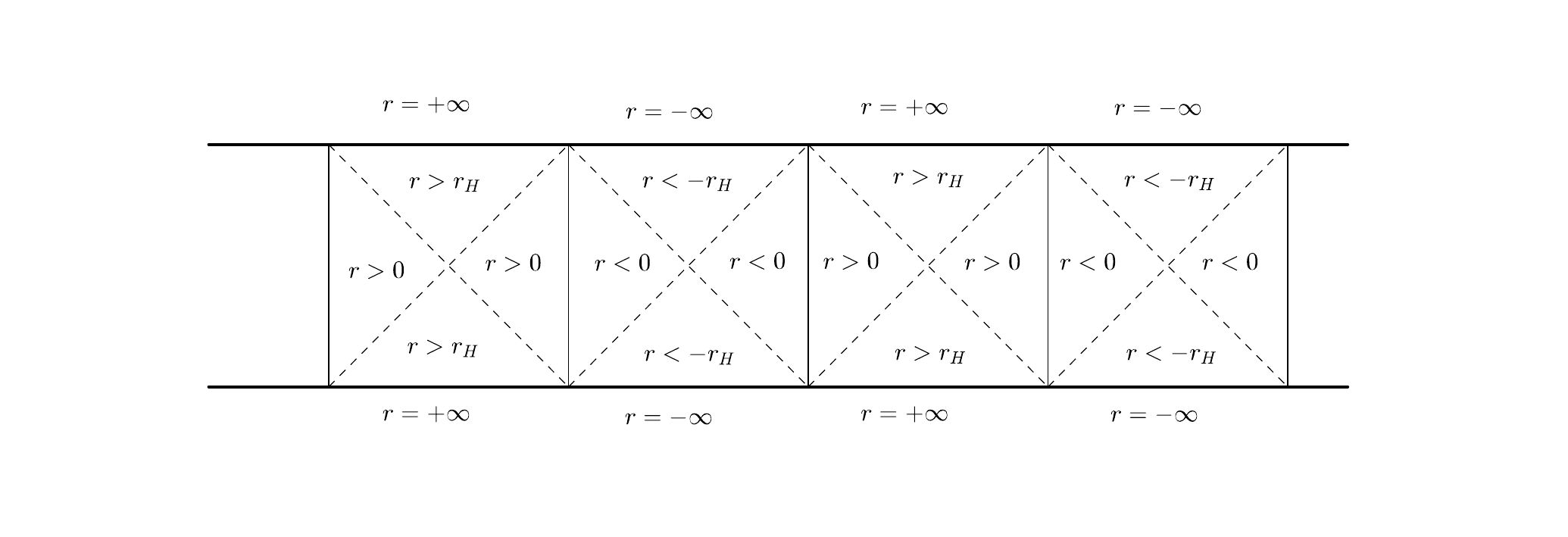}}
\hss}
\caption{Conformal diagram of wormhole with $\Lambda>0$. 
}
\label{Fg4}
\end{figure}
The global structure of the $\Lambda>0$ wormhole is more complex.
The wormhole then connects through the disk at $r=0$ de Sitter regions with $r>0$ 
and with cosmological horizon 
at $r=r_H$ to those with $r<0$ and with  cosmological horizons at $r=-r_H$. 
As shown in Fig.\ref{Fg4}, this gives rise to an infinite sequence of alternating $r>0$ and $r<0$ regions. 
This diagram can be obtained by considering the Kruskal extension, however, one can apply 
a much simpler method just to see that the diagram is periodic. 

Restricting  to the region(s) where $r^2\leq 3/\Lambda$ and setting 
\be
r=\sqrt{\frac{3}{\Lambda}}\,\sin\chi,~~~~~~t=\sqrt{\frac{3}{\Lambda}}\,\tilde{t},~~~~~~\zeta^2=\frac{\Lambda a^2}{3},
\ee
the metric \eqref{00} becomes
\bea                     \label{04}
ds^2=\frac{3}{\Lambda}\left(
-\frac{1+\zeta^2\cos^2\vartheta}{1+\zeta^2}\,\cos^2\chi\,d\tilde{t}^2
+\frac{\sin^2\chi+\zeta^2\cos\vartheta}{\sin^2\chi+\zeta^2}\,d\chi^2  \right. \nonumber  \\
+\left.\frac{\sin^2\chi+\zeta^2\cos^2\vartheta}{1+\zeta^2\cos^2\vartheta}\,d\vartheta^2
+\frac{\sin^2\chi+\zeta^2}{1+\zeta^2}\,\sin^2\vartheta d\varphi^2
\right).
\eea
The wormhole throat is placed where $\sin\chi=0$ while the cosmological horizon is at 
$\cos\chi=0$. 
The advantage of this parameterization is that the range of the radial coordinate $\chi$ 
can be analytically extended to the whole line $(-\infty,\infty)$, the metric coefficients then becoming
periodic functions 
of $\chi$. This explains the periodicity of the spacetime diagram in Fig.\ref{Fg4} containing 
an infinite sequence of wormhole throats at $\chi=\pi k$ and ``Einstein-Rosen bridges" at  $\chi=\pi(k+1/2)$. 
However, the coordinates used in \eqref{04} are not global and cover only the interior of the diamonds 
in Fig.\ref{Fg4}. 

\subsection{Separation of variables in the wave equation}

As a last remark, returning to \eqref{00}, we notice that 
coordinates used in this metric allow one to separate the  variables in the Klein-Gordon equation 
\be                          \label{KG}
\left(\Box-\mu^2\right)\Phi=0.
\ee
{For $\Lambda\to 0$ this is not surprising since the variables  in the wave equation 
  (and in the the Hamilton-Jacobi equation  \cite{Pretorius:1998sf}) separate
 in flat space
 expressed in the spheroidal coordinates.} 
 The variables separate also for $\Lambda\neq 0$ 
 if the metric is expressed in  Boyer-Lindquist coordinates \cite{Carter:1968ks,Chong:2004hw},
 but  it is not immediately obvious that they separate  in spheroidal coordinates. 
 However, 
 setting $\Phi=F(r)G(\vartheta)\exp\{i\omega t+i m\varphi\}$ we find that 
 \eqref{KG} reduces to 
  \bea
 \left((r^2+a^2)D\,F^\prime(r)\right)^\prime+\left(\frac{3\Xi\omega^2 }{\Lambda D}-\mu^2 r^2
 +\frac{m^2a^2\Xi }{r^2+a^2  }+\lambda
 \right)F(r)=0, \nonumber \\
 \left.\left.\frac{1}{\sin\vartheta}\right(\sin\vartheta\, \Delta_\vartheta\,G^\prime(\vartheta)\right)^\prime 
 -\left(\frac{3\Xi\omega^2 }{\Lambda \Delta_\vartheta}+
 \mu^2 a^2\cos^2\vartheta+\frac{m^2 \Xi}{\sin^2\vartheta}+\lambda
 \right)G(\vartheta)=0,
 \eea
where $\lambda$ is the separation constant determined by the condition of regularity of $G(\vartheta)$.  

\section{Conclusions}
To summarize, we have shown that the zero mass limit of the Kerr spacetime  is not  flat Minkowski 
space as is usually assumed 
 but a locally flat static  wormhole spacetime containing a conical singularity of the Ricci tensor along a ring. 
 This singularity can be interpreted as an effect of a singular matter source -- a negative tension 
 cosmic string loop. Similarly, the zero mass limit of the Kerr-AdS or Kerr-de-Sitter spacetime is 
 a locally (A)dS wormhole supported by a negative tension ring. This yields probably the simplest  way 
 to construct wormholes -- by taking limits of the known metrics. 
 
 As a final remark, we notice that the Kerr spacetime can be ``mutilated" and restricted to the $r\geq 0$ 
 region by introducing an additional matter source distributed over the disk encircled by the ring 
 singularity \cite{Israel:1970kp}. The $M\to 0$ limit of such ``mutilated" spacetime 
 would be flat Minkowski space. 
The $M\to 0$ limit of the full Kerr spacetime is the wormhole.

\section*{Acknowledgements}
We are indebted to G\'erard Cl\'ement for his remark about similarity of our wormhole metric \eqref{0}
with the  zero mass  Kerr metric, which  initiated our study. 
G.W.G. thanks the LMPT for hospitality and acknowledges the support of  ``Le Studium" -- Institute
for Advanced Studies of the Loire Valley. 
M.S.V. was partly supported by the Russian Government Program of Competitive Growth 
of the Kazan Federal University.

%\bibliographystyle{apsrev4-1}
%\bibliography{/Users/mvolkov/Dropbox/TEX/W}


\end{document}